\documentclass[11pt]{article}

\pdfoutput=1

\usepackage[makeroom]{cancel}
\usepackage{color}
\usepackage{graphicx}
\usepackage{amsmath}
\usepackage{amssymb}
\usepackage{xspace}
\usepackage[small]{subfigure}
\usepackage[numbers,compress]{natbib}
\usepackage[hyperfootnotes=false]{hyperref}

\newlength{\fighskip} \fighskip=2pt
\newlength{\figvskip} \figvskip=3pt

\newcommand*{\figbox}[2]{{
  \def\figscale{#1}
  \def\arraystretch{0.8}
  \arraycolsep=0pt
  \begin{array}{c}
    \vbox{\vskip\figscale\figvskip
      \hbox{\hskip\figscale\fighskip
        \includegraphics[scale=\figscale]{#2}}}
  \end{array}}}

\usepackage{mciteplus} 
\usepackage{dcolumn}
\usepackage{bm}
\usepackage{verbatim}
\usepackage{amscd}
\usepackage{amsfonts}
\usepackage{setspace}
\usepackage{amsthm}
\usepackage{enumerate}
\usepackage{mathtools}

\theoremstyle{plain}

\theoremstyle{plain}

\theoremstyle{plain}
 
\theoremstyle{plain}

\theoremstyle{remark}

\theoremstyle{conjecture}

\theoremstyle{observation}

\theoremstyle{definition}

\theoremstyle{corollary}

\theoremstyle{definition}

\theoremstyle{definition}

\theoremstyle{result}

\theoremstyle{assumption}

\theoremstyle{definition}

\theoremstyle{problem}

\theoremstyle{fact}

\DeclareMathOperator{\Tr}{Tr}

\begin{document}

\title{\bf Soft mode and interior operator in \\
Hayden-Preskill thought experiment}
\author{
Beni Yoshida\\ 
{\em \small Perimeter Institute for Theoretical Physics, Waterloo, Ontario N2L 2Y5, Canada} }
\date{}

\maketitle

\begin{abstract}
We study the smoothness of the black hole horizon in the Hayden-Preskill thought experiment by using two particular toy models based on variants of Haar random unitary. The first toy model corresponds to the case where the coarse-grained entropy of a black hole is larger than its entanglement entropy. We find that, while the outgoing mode and the remaining black hole are entangled, the Hayden-Preskill recovery cannot be performed. The second toy model corresponds to the case where the system consists of low energy soft modes and high energy heavy modes. We find that the Hayden-Preskill recovery protocol can be carried out via soft modes whereas heavy modes give rise to classical correlations between the outgoing mode and the remaining black hole. We also point out that the procedure of constructing the interior partners of the outgoing soft mode operators can be interpreted as the Hayden-Preskill recovery, and as such, the known recovery protocol enables us to explicitly write down the interior operators. Hence, while the infalling mode needs to be described jointly by the remaining black hole and the early radiation in our toy model, adding a few extra qubits from the early radiation is sufficient to reconstruct the interior operators. 
\end{abstract}

\section{Introduction}

Almost forty years since its formulation, the black hole information problem and its variants still shed new lights on deep conceptual puzzles in quantum gravity, and also provides useful insights to study strongly interacting quantum many-body systems~\cite{Hawking75}. While the ultimate solution of the problem could be obtained only by experimental observations, progresses can be made by utilizing thought experiments based on simple toy models~\cite{Page93, Page:1993aa}. In the last decade, two particular thought experiments on black hole dynamics have fascinated and puzzled theorists; the Almheiri-Marolf-Polchinski-Sully (AMPS) thought experiment~\cite{Almheiri13} and the Hayden-Preskill thought experiment~\cite{Hayden07}. The AMPS thought experiment suggests that the smooth horizon in an old black hole, which is a consequence of the equivalence principle, may be inconsistent with monogamy of entanglement~\footnote{A related argument has appeared in~\cite{Braunstein:2013aa}}. The Hayden-Preskill thought experiment poses questions concerning the absoluteness of the event horizon by suggesting that an object which has fallen into a black hole may be recovered. While there have been refined arguments and counterarguments on these conclusions in more realistic physical settings, essential features of the original works can be reduced to very simple calculations based on Haar random unitary operators. 

The main idea of this paper centers around a tension between the smooth horizon and the Hayden-Preskill thought experiment. The no-firewall postulate asserts the presence of entanglement between the infalling and outgoing Hawking pair. The recoverability in the Hayden-Preskill thought experiment requires the outgoing Hawking radiation to be entangled with a joint system of the early radiation and the reference qubits of the infalling quantum state. However, due to monogamy of entanglement, the no-firewall postulate and the recoverability look mutually incompatible. 

Here we seek for a resolution to this tension by arguing that the ``outgoing Hawking radiation'' in the Hayden-Preskill and AMPS experiments are actually different degrees of freedom. While it would be desirable to demonstrate such a separation of the Hilbert space of the outgoing mode via direct calculations on actual models of quantum gravity, our goal is more modest. In this paper, we will study a certain refinement of Haar random unitary dynamics. The unitary operator $U$ preserves the total global $U(1)$ charge and acts as Haar random unitary operator in each subspace with fixed charge in a block diagonal manner. Although charges in black holes has led to intriguing puzzles in quantum gravity~\cite{Arkani-Hamed:2007aa}, it is not our primary goal to study the effect of global charges on the AMPS and Hayden-Preskill thought experiments. Our primary focus is on unitary dynamics which preserves energy. We utilize the block diagonal structure of $U(1)$-symmetric Haar random unitary to capture ergodic dynamics which essentially acts as Haar random unitary on each small energy window. This is partly motivated from recent works where $U(1)$-symmetric local random unitary circuits successfully capture key properties of energy conserving systems such as an interplay of diffusive transport phenomena and the ballistic operator growth~\cite{Khemani:2018aa, Rakovszky:2018aa}. 

We will show that $U(1)$-symmetric modes are responsible for the Hayden-Preskill recovery whereas the non-symmetric modes are responsible for correlations between the infalling and the outgoing Hawking pair. With an actual physical system with energy conservation in mind, we interpret the symmetric and non-symmetric modes as low energy (soft) and high energy (heavy) modes respectively. Namely we claim that the Hayden-Preskill thought experiment can be carried out by using soft modes which are distinct from the Hawking radiation. Such low energy modes may be the pseudo Goldstone mode which corresponds to the 't Hooft's gravitational mode~\cite{Hooft:1987aa}. Or perhaps they may correspond to soft gravitons due to spontaneous breaking of supertranslation symmetries~\cite{Strominger:2014aa}. In our toy model, however, the correlation between the infalling and outgoing Hawking modes is found to be purely \emph{classical} as opposed to quantum correlations in the Hawking pair which would be seen by an infalling observer. Namely, the outgoing soft mode is found entangled with a joint of the remaining black hole and the early radiation, not with the remaining black hole itself.

We also discuss the construction of the interior partner operators of the outgoing Hawking mode. While the partners of the outgoing soft mode operators cannot be found in the remaining black hole, adding a few qubits from the early radiation to the remaining black hole is enough to construct the interior operators. The key observation is that the reconstruction of the interior operators can be seen as the Hayden-Preskill thought experiment, and hence the method from~\cite{Yoshida:2017aa} can be used to explicitly write down the interior operators. This observation enables us to show that the black hole interior modes are robust against perturbations on the early radiation due to scrambling dynamics. 

From the perspective of the AMPS puzzle, our result appears to suggest that interior soft operators need to be reconstructed in the early radiation instead of the remaining black hole. While this resonates with previous approaches bundled under ``$A=R_{B}$'' or ``$\text{ER} = \text{EPR}$''~\cite{Almheiri13b, Papadodimas:2013aa, Maldacena13, Susskind13, Bousso:2013ab}, these run into various paradoxes~\cite{Marolf:2013aa, Bousso:2013aa, Giddings:2013aa} (See~\cite{Almheiri13b, Harlow:2016ab} for summaries) as the construction may lead to apparent non-local encoding between the interior and the early radiation. Relatedly, construction of interior operators is state-dependent, which may suffer from a number of potential inconsistencies with quantum mechanics. For further discussion, please see a selection (but by all means not a complete set) of recent work~\cite{Marolf:2013aa, Bousso:2014aa, Bousso:2013aa, Bousso:2013ab, Marolf:2016aa, Harlow:2014aa, Papadodimas:2013aa, Papadodimas:2014aa, Papadodimas:2014ab, Raju:2017aa, Jafferis17}. 
In an accompanying paper, we will debug these problems and present constructions of interior operators which are local (\emph{i.e.} without involving the early radiation) and state-independent (\emph{i.e.} no dependence on the initial state of the black hole) by incorporating the effect of backreaction by the infalling observer explicitly~\cite{Yoshida:2019}.

The paper is organized as follows. In section~\ref{sec:subspace}, we warm up by studying the case where the black hole is entangled only through a subspace and its evolution is given by Haar random unitary. A corresponding physical situation is that the entanglement entropy $S_{E}$ of a black hole is smaller than its coarse-grained entropy $S_{\text{BH}}$. The eternal AdS black hole corresponds to $S_{E} = S_{\text{BH}}$ whereas a one-sided pure state black hole corresponds to $S_{E}=0$. We will see that taking $S_{E} < S_{\text{BH}}$ generates quantum entanglement between the outgoing mode and the remaining black hole, but the Hayden-Preskill recovery is no longer possible, highlighting their complementary nature. In section~\ref{sec:symmetric}, we analyze the case where the black hole is entangled through a $U(1)$-symmetric subspace and its evolution is given by $U(1)$-symmetric Haar random unitary. Physically this corresponds to a black hole which is entangled with its partner through the subspace consisting of typical energy states at given temperature. In section~\ref{sec:recovery}, we present concrete recovery protocols by following \cite{Yoshida:2017aa}. In section~\ref{sec:mirror}, we describe the procedure to construct the interior operators of the outgoing soft mode. In section~\ref{sec:discussion}, we conclude with discussions. 

\begin{figure}
\centering
\includegraphics[width=0.35\textwidth]{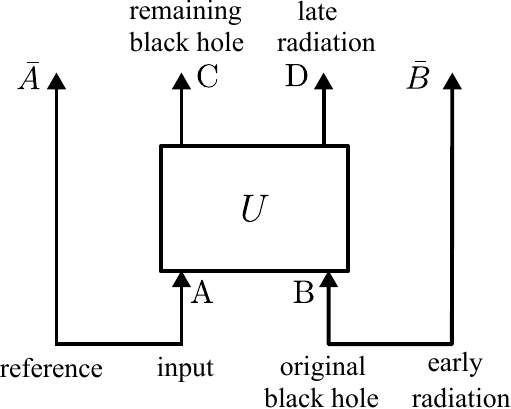} 
\caption{The Hilbert space structure. 
\label{fig-HP-notation}
}
\end{figure}

Before delving into detailed discussions, we establish a few notations used throughout this paper. See Fig.~\ref{fig-HP-notation}. We will denote the Hilbert spaces for the input quantum state as $A$, the original black hole as $B$, the remaining black hole as $C$ and the late Hawking radiation as $D$. It is convenient to introduce the reference Hilbert space for the input quantum state. See~\cite{Hayden07, Hosur:2015ylk} for detailed discussions on the use of the reference system. The reference Hilbert space is denoted by $\bar{A}$. The entangled partner of $B$ is denoted by $\bar{B}$. The unitary dynamics $U$ of a black hole acts on $AB \simeq CD$. The Hilbert space dimension of a subsystem $R$ is denoted by $d_{R}$ while the number of qubits on $R$ is denoted by $n_{R}$. Entropies are computed as binary entropies. 

The Hayden-Preskill thought experiment with Haar random unitary with various global symmetries was studied independently by Nakata, Wakakuwa and Koashi. They pointed out that, for $U(1)$ symmetry with generic input states, the recovery requires collecting extensive number of qubits. A similar conclusion is obtained in appendix~\ref{appendix} for non-symmetric input states. Upon completion of this work, we became aware of an independent work~\cite{Nomura18} which addresses the black hole evaporation process with distinction between hard and soft modes. 

\subsection{Summary of diagrammatic techniques}

In this paper, we will extensively use diagrammatic tensor notations in order to express wavefunctions and operators as well as physical processes. Here we provide a brief tour of key properties for readers who are not familiar with these techniques. 

Wavefunctions and operators are represented by 
\begin{align}
|\psi\rangle \ = \ \figbox{1.0}{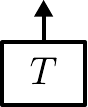} \qquad \langle \psi| = \figbox{1.0}{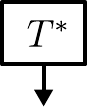}
\qquad O \ =\ \figbox{1.0}{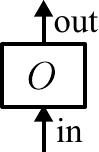}
\end{align}
which can be also explicitly written as
\begin{align}
|\psi\rangle = \sum_{j} T_{j}|j\rangle \qquad \langle \psi |\ = \ \sum_{j} T_{j}^* \langle j | \qquad O\ = \  \sum_{ij} O_{ij}|i\rangle \langle j |.
\end{align}
By using these tensors as building blocks, one can express various physics in a graphical manner. For instance, an expectation value can be represented by
\begin{align}
\langle \psi | O_{n}\cdots O_{1} |\psi\rangle \ =\ \figbox{1.0}{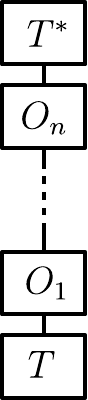}\ .
\end{align}
In order to associate a physical process to an equation like $\langle \psi | O_{n}\cdots O_{1} |\psi\rangle$, one needs to read it from the right to the left, \emph{i.e.} the initial state $|\psi\rangle$ is acted by $O_{1},O_{2},\cdots$ sequentially and then is projected onto $|\psi\rangle$. In the diagrammatic notation, one needs to read the figure from the bottom to the top, \emph{i.e.} the time flows upward in the diagram.

A key (yet sometimes confusing) feature of tensor diagrams is that the same tensor can represent different physical processes depending on which tensor indices are used as inputs and outputs. Let us look at a few important examples. An identity operator, $I = \sum_{j} |j\rangle \langle j |$, can be expressed as a straight line (\emph{i.e.} a trivial tensor) since its inputs and outputs are the same:
\begin{align}
I \ = \ \figbox{1.0}{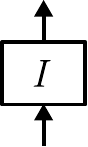}  \ = \ \figbox{1.0}{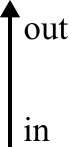}
\end{align}
This diagram has one input leg (index) and one output leg. One may bend the line and construct the following diagram:
\begin{align}
|\text{EPR}\rangle \propto \sum_{j} |j\rangle \otimes |j\rangle  \ = \ \figbox{1.0}{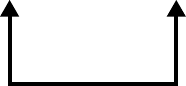}
\end{align}
which is the same trivial line, but with two output legs instead of one in and one out. This diagram represents an unnormalized EPR pair defined on $\mathcal{H}^{\otimes 2}$.

Another important example involves a transpose of an operator:
\begin{align}
O \ =\ \figbox{1.0}{fig-notation-O} \ =\ \figbox{1.0}{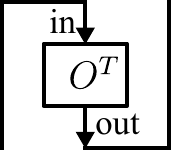}
\end{align}
where
\begin{align}
O = \sum_{i,j} O_{ij}|i\rangle \langle j| \qquad O^{T} = \sum_{i,j} O_{ji}|i\rangle \langle j|. 
\end{align}
Here the transpose $O^{T}$ exchanges the input and output of $O$. The original diagram with $O$ represents a physical process where an arbitrary input wavefunction $|\psi\rangle$ is acted by $O$ and $O|\psi\rangle$ appears as an output. The second diagram with $O^{T}$ describes a physical process which involves three Hilbert spaces of the same size $\mathcal{H}^{\otimes 3} = \mathcal{H}_{1}\otimes \mathcal{H}_{2} \otimes \mathcal{H}_{3}$:
\begin{align}
\figbox{1.0}{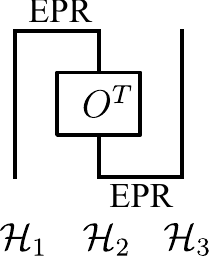}
\end{align}
Here $\mathcal{H}_{1}$ supports an arbitrary input wavefunction whereas $\mathcal{H}_{2}\otimes \mathcal{H}_{3}$ starts with the EPR pair. Then, the transpose $O^{T}$ acts on $\mathcal{H}_{2}$, and then the system is projected onto the EPR pair on $\mathcal{H}_{1}\otimes \mathcal{H}_{2}$. The outcome on $\mathcal{H}_{3}$ is $O |\psi\rangle$. One may represent this explicitly as the following equation:
\begin{align}
(\langle \text{EPR}|_{12} \otimes I_{3} )(I_{1} \otimes O^T_{2} \otimes I_{3})|\psi\rangle_{1} \otimes |\text{EPR}\rangle_{23} \propto O |\psi\rangle_{3}.
\end{align}
As the above examples suggest, one may interpret an operator as a quantum state and vice versa. To assign physical interpretations to the diagram, we simply read it from the bottom to the top.

\section{Hayden-Preskill with code subspaces}\label{sec:subspace}

In this section, we revisit the Hayden-Preskill thought experiment and the AMPS problem for quantum black holes which are entangled through ``code subspaces''. 

Following the previous works, we will model a black hole as an $n$-qubit quantum system with $n=S_{\text{BH}}$ where $S_{\text{BH}}$ is the coarse-grained entropy which is proportional to the area of the black hole. 
We will also assume that the dynamics of a black hole is given by a Haar random unitary operator.
Let $S_{\text{E}}$ be the entanglement entropy between the black hole and all the other degrees of freedom including the early radiation. 
Previous discussions on the Hayden-Preskill thought experiment and the AMPS thought experiment concern maximally entangled black holes with $S_{\text{BH}}=S_{\text{E}}$. 
Here we will investigate a black hole with $S_{\text{BH}}>S_{\text{E}}$. 
Black holes before the Page time satisfy this condition.
Similar situations have been previously considered by Verlinde and Verlinde~\cite{Verlinde:2013aa}. 
In particular, they pointed out that taking $S_{\text{BH}}>S_{\text{E}}$ enables us to construct partner operators of the outgoing mode in the degrees of freedom of the remaining black hole. 
Furthermore, they proposed a scenario to resolve the AMPS puzzle by using this effect.

The aim of this section is to study the consequence of taking $S_{\text{BH}}>S_{\text{E}}$ in the Hayden-Preskill thought experiment. We find that, if a black hole with $S_{\text{BH}} \gg S_{\text{E}}$ evolves under Haar random unitary operator, the Hayden-Preskill recovery cannot be performed unless one collects $O(n)$ qubits from the outgoing Hawking radiation. In fact, we find that the Hayden-Preskill recoverability and the smoothness of the horizon are mutually incompatible phenomena within the applicability of toy descriptions based on Haar random untary operator.

\subsection{Haar integral}

The quantum state we are interested in is the following with $d_{\bar{A}}\leq d_{A}$ and $d_{\bar{B}}\leq d_{B}$:
\begin{align}
|\Psi\rangle = \ \figbox{1.0}{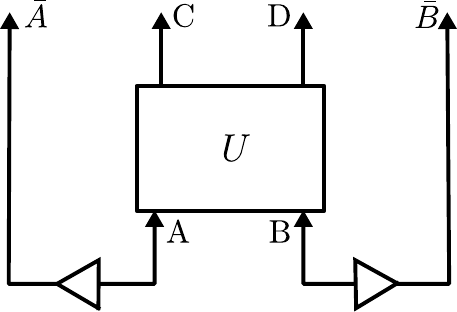}\ = \frac{1}{\sqrt{d_{\bar{A}}d_{\bar{B}}}} U_{k\ell m o} |k \rangle_{\bar{A}} |\ell \rangle_{\bar{B}} | m \rangle_{C} | o \rangle_{D} \label{eq:world}
\end{align}
where summations are implicit with $k=1,\ldots d_{\bar{A}}$, $\ell = 1, \ldots, d_{\bar{B}}$, $m = 1,\ldots, d_{C}$ and $o = 1,\ldots, d_{D}$. Triangles represent normalized isometries. For instance, the input state in $\bar{A}A$ is given by
\begin{align}
\frac{1}{\sqrt{d_{\bar{A}}}} \sum_{k=1}^{d_{\bar{A}}} |k \rangle_{\bar{A}} |k\rangle_{A}.
\end{align}
Here $|k\rangle_{A}$ spans only a subspace of $A$. The choice of $|k\rangle_{A}$ is not important as the system evolves by Haar random unitary. 

Each subsystem, $\bar{A}AB\bar{B}CD$, admits the following physical interpretation in the Hayden-Preskill thought experiment. $A$ and $\bar{A}$ correspond to Hilbert spaces for the input quantum state and its reference system respectively. An input quantum information is drawn from $d_{\bar{A}}$-dimensional subspace $\bar{A}$ and is encoded into $d_{A}$-dimensional subspace $A$ before thrown into a black hole.  $B$ corresponds to the coarse-grained Hilbert space of a black hole whereas $\bar{B}$ corresponds to the subspace where the black hole is entangled. The initial black hole satisfies $S_{\text{BH}}>S_{\text{E}}$ since
\begin{align}
S_{\text{BH}} = \log d_{B} \qquad S_{\text{E}} = \log d_{\bar{B}}. 
\end{align}
Finally, $C$ and $D$ corresponds to the remaining black hole and the outgoing Hawking radiation respectively. 

The Haar average formula with two $U$s and two $U^{\dagger}$s is
\begin{align}
\int dU \ U_{i_1 j_1} U_{i_2 j_2} U^{*}_{i_1' j_1'} U^{*}_{i_2' j_2'} = &\ \frac{1}{d^2 -1 } \left( \delta_{i_1 i_1' }\delta_{i_2 i_2'} \delta_{j_1 j_2'} \delta_{j_2 j_1'} +  \delta_{i_1 i_2' }\delta_{i_2 i_1'} \delta_{j_1 j_1'} \delta_{j_2 j_2'}   \right) \\ 
 &- \frac{1}{d(d^2 -1 )}  \left( \delta_{i_1 i_1' }\delta_{i_2 i_2'} \delta_{j_1 j_2'} \delta_{j_2 j_1'} +  \delta_{i_1 i_2' }\delta_{i_2 i_1'} \delta_{j_1 j_1'} \delta_{j_2 j_2'} \right)
\end{align}
where $d=2^n$. Approximating $d^2-1 \approx d^2$, this lets us compute the Haar average of $\Tr\{ \rho_{C}^2 \}$ and $\Tr\{ \rho_{\bar{B}D}^2 \}$:
\begin{align}
&\int dU \ \Tr \{ \rho_{C}^2 \} = \frac{1}{d_{\bar{A}}^2 d_{\bar{B}}^2} \int dU\ U_{k\ell m o} U^{*}_{k \ell m' o} U_{k' \ell' m' o'} U^{*}_{k' \ell' m o'} \\
&\int dU \  \Tr \{ \rho_{\bar{B}D}^2 \} = \frac{1}{d_{\bar{A}}^2 d_{\bar{B}}^2} \int dU\ U_{k\ell m o} U^{*}_{k \ell' m o'} U_{k' \ell' m' o'} U^{*}_{k' \ell m' o},
\end{align}
and after simple calculations of delta functions, we find
\begin{align}
&\int dU\ \Tr \{ \rho_{C}^2 \} = \frac{1}{d_{C}} + \frac{1}{d_{D} d_{\bar{A}}d_{\bar{B}}} - \frac{d_{C}}{d^2} - \frac{1}{d d_{C}d_{\bar{A}}d_{\bar{B}} } \approx  \frac{1}{d_{C}} + \frac{1}{d_{D} d_{\bar{A}}d_{\bar{B}}} \\
&\int dU\ \Tr \{ \rho_{\bar{B}D}^2 \} = \frac{1}{d_{D}d_{\bar{B}}} + \frac{1}{d_{C}d_{\bar{A}}}  - \frac{1}{dd_{C}d_{\bar{A}}} - \frac{1}{d d_{D}d_{\bar{A}}} \approx \frac{1}{d_{D}d_{\bar{B}}} + \frac{1}{d_{C}d_{\bar{A}}}
\end{align}
where terms with $1/d$ factors are ignored.

\subsection{Recoverability}
  
The Hayden-Preskill thought experiment concerns recoverability of the input quantum state on $\bar{A}$ by having an access to both the early radiation $\bar{B}$ and the outgoing mode $D$. As such, the recoverability of a quantum state can be studied by asking whether one can distill EPR pairs between $\bar{A}$ and $D\bar{B}$. One can analyze the recoverability by studying quantum correlations between $\bar{A}$ and $\bar{B}D$ in $|\Psi\rangle$ in Eq.~\eqref{eq:world}. Hence, as a measure of recoverability, we will use the R\'{e}nyi-$2$ mutual information, defined by $I^{(2)}(\bar{A},\bar{B}D)=S^{(2)}_{\bar{A}}+S^{(2)}_{\bar{B}D} - S^{(2)}_{\bar{A}\bar{B}D}$ by following~\cite{Hosur:2015ylk}~\footnote{Strictly speaking, the R\'{e}nyi-$2$ mutual information is not an entanglement monotone in general. See~\cite{Yoshida:2019aa} for details.}.

We can compute the R\'{e}nyi-$2$ mutual information between $\bar{A}$ and $\bar{B}D$:
\begin{align}
\int dU \ 2^{I^{(2)}(\bar{A},\bar{B}D)} \approx d_{\bar{A}} \frac{d_{D}d_{\bar{A}}d_{\bar{B} }+ d_{C}}{d_{D}d_{\bar{B}} + d_{C}d_{\bar{A}} }.
\end{align}
Let us look at recoverability in three regimes. 

\begin{enumerate}[(a)]
\item (small $d_{D}$) For $d \frac{1}{d_{\bar{A}} d_{\bar{B}}} \gg d_{D}^2$, we find
\begin{align}
d_{D}d_{\bar{A}}d_{\bar{B} } \ll d_{C}, \quad d_{D}d_{\bar{B}} \ll d_{C}d_{\bar{A}}
\end{align}
leading to
\begin{align}
2^{I^{(2)}(\bar{A},\bar{B}D)}  \approx 1,
\end{align}
implying that $\bar{A}$ and $\bar{B}D$ are not correlated. 
\item (intermediate $d_{D}$) For $d \frac{d_{\bar{A}}}{d_{\bar{B}}} \gg d_{D}^2 \gg d \frac{1}{d_{\bar{A}} d_{\bar{B}}} $, we find 
\begin{align}
d_{D}d_{\bar{A}}d_{\bar{B} } \gg d_{C}, \quad d_{D}d_{\bar{B}} \ll d_{C}d_{\bar{A}}
\end{align}
leading to
\begin{align}
2^{I^{(2)}(\bar{A},\bar{B}D)}  \approx \frac{d_{D}^2 d_{\bar{A}}d_{\bar{B}}}{d}.
\end{align}
The mutual information increases by two as the number of qubits in $D$ increases by one. In order for this regime to be present, we need $d_{\bar{A}}^2 \gg 1$. 
\item (large $d_{D}$) For $d_{D}^2 \gg d \frac{d_{\bar{A}}}{d_{\bar{B}}}$, we find 
\begin{align}
d_{D}d_{\bar{A}}d_{\bar{B} } \gg d_{C}, \quad d_{D}d_{\bar{B}} \gg d_{C}d_{\bar{A}}
\end{align}
\begin{align}
2^{I^{(2)}(\bar{A},\bar{B}D)}  \approx d_{\bar{A}}^2,
\end{align}
implying that the correlation is nearly maximal. 
\end{enumerate}

\subsection{Smoothness}

Next let us turn our attention to the AMPS problem. Here we are interested in whether partners of outgoing mode operators on $D$ can be reconstructed on the remaining black hole $C$ or not. This is possible if and only if $C$ and $D$ retain strong quantum correlations. Hence, as a measure of smoothness of the horizon, we will use $I^{(2)}(C,D)$. 

We compute the Haar average of the mutual information between $C$ and $D$:
\begin{align}
\int dU \ 2^{I^{(2)}(C,D)} \approx \frac{d_{C} d_{D}^2 }{d_{D}d_{\bar{A}}d_{\bar{B}} + d_{C} }.
\end{align}
We also derive the mutual information between $\bar{A}\bar{B}$ and $D$ in order to illustrate the monogamy of entanglement. The two mutual information are related as follows
\begin{align}
2^{I^{(2)}(\bar{A}\bar{B},D)} = \frac{2^{S^{(2)}_{D}}}{2^{I^{(2)}(C,D)}} \approx \frac{d_{D}^2}{2^{I^{(2)}(C,D)}}
\end{align}
where we approximated $\rho_{D}$ by a maximally mixed state.

\begin{enumerate}[(a)]
\item (small $d_{D}$) For $d\frac{1}{d_{\bar{A}} d_{\bar{B}}} \gg d_{D}^2$, we find
\begin{align}
2^{I^{(2)}(C,D)}  \approx  d_{D}^2, \qquad 2^{I^{(2)}(\bar{A}\bar{B},D)}  \approx 1
\end{align}
implying that $I^{(2)}(\bar{A}\bar{B},D)\approx 0$ and does not increase as $d_{D}$ increases. 
\item[(bc)] (intermediate and large $d_{D}$) For $d_{D}^2 \gg  d \frac{1}{d_{\bar{A}} d_{\bar{B}}} $, we find
\begin{align}
2^{I^{(2)}(C,D)}  \approx  \frac{d}{d_{\bar{A}} d_{\bar{B}}}, \qquad 2^{I^{(2)}(\bar{A}\bar{B},D)}  \approx \frac{d_{D}^2  d_{\bar{A}} d_{\bar{B}} }{ d }
\end{align}
implying that $I^{(2)}(\bar{A}\bar{B},D)$ becomes large as the number of qubits in $D$ increases while $I^{(2)}(C,D)$ remains unchanged. 
\end{enumerate}

\subsection{Physical interpretation}

Recall that the initial black hole satisfies $S_{\text{BH}}= \log d_{B}$ and $S_{E} = \log d_{\bar{B}}$ in our toy model. For the recovery in the Hayden-Preskill thought experiment, we need 
\begin{align}
d_{D} \geq \sqrt{d_{A}d_{\bar{A}}}\sqrt{\frac{d_{B}}{d_{\bar{B}}}}.
\end{align}
When a black hole is maximally entangled with $d_{B}= d_{\bar{B}}$, the above lower bound reduces to 
\begin{align}
d_{D} \geq \sqrt{d_{A}d_{\bar{A}}}
\end{align}
reproducing the result in~\cite{Yoshida:2017aa}.

Here we are interested in cases with $d_{B}\gg d_{\bar{B}}$ where the coarse-grained entropy is larger than the entanglement entropy. Namely, if the number of qubits in $B$ and $\bar{B}$ satisfy $n_{B} - n_{\bar{B}}\sim O(n)$, then Bob needs to collect an extensive number of qubits. As such, if $d_{B}\gg d_{\bar{B}}$, simple recovery is not possible (unless one collects an extensive number of qubits). If we take $d_{\bar{B}}=1$, we have $d_{D} \geq d_{A} \sqrt{d_{B}} \sim O(\sqrt{d})$ implying that Bob needs to collect more than a half of the total qubits for reconstruction as pointed out by Page~\cite{Page93}.

The failure of recovery for $d_{B} \gg d_{\bar{B}}$ can be understood from the calculation of $I^{(2)}(C,D)$. When $d_{D}$ is small, most of qubits in $D$ are entangled with the remaining black hole $C$ and do not reveal any information about $\bar{A}\bar{B}$. Once $2^{I^{(2)}(C,D)}$ reaches a stationary value of $\frac{d}{(d_{\bar{A}} d_{\bar{B}})^2}$, entanglement between $D$ and $\bar{A}\bar{B}$ starts to develop. From the perspective of the firewall puzzle, the increase of $I^{(2)}(C,D)$ appears to suggest that partner operators of the outgoing mode $D$ can be found on the remaining black hole $C$. Verlinde and Verlinde employed this mechanism as a possible resolution of the firewall puzzle and made an intriguing relation to theory of quantum error-correction~\cite{Verlinde:2013aa}. These observations illustrate that Hayden-Preskill and the smoothness of the horizon are mutually complementary phenomena when $U$ is a Haar random unitary which throughly mixes the Hilbert space $AB$. Namely, the presence of partner operators in $C$ requires that $D$ is correlated with $C$ whereas the recoverability of an input quantum state in the Hayden-Preskill thought experiment requires that $D$ is entangled with $\bar{A}\bar{B}$. In fact, the tradeoff between the smoothness and recoverability is strikingly sharp; once $D$ becomes large enough to start releasing information about $\bar{A}$ via increase of $I^{(2)}(\bar{A},\bar{B}D)$, the growth of $I^{(2)}(C,D)$ stops. 

Finally, we make a comment on the complexity of performing recovery protocols in the Hayden-Preskill thought experiment. For $d_{B}=d_{\bar{B}}$, a simple recovery protocol is known to exist~\cite{Yoshida:2017aa}. When applying this method to the case with large $d_{D}$, it is crucial to identify degrees of freedom in $D$ which is not entangled with $C$. Under chaotic dynamics of a black hole, it is plausible to expect that such degrees of freedom become non-local inside $D$ and require complex operations. Hence, performing the Hayden-Preskill recovery may be unphysical when $d_{B}\gg d_{\bar{B}}$. In fact, it may be more correct to say that $D$ corresponds to simple degrees of freedom which can be accessed easily from the outside whereas $C$ corresponds to complex ones.

\section{Hayden-Preskill with $U(1)$ symmetry}\label{sec:symmetric}

In this section, we study the Hayden-Preskill thought experiment in the presence of conserved quantities. For simplicity of discussions, we will consider systems with $U(1)$ global symmetry where the total spin in the $z$-direction is preserved. We will find that the Hayden-Preskill recovery is possible only if an input quantum state is embedded into a subspace with fixed charges.

\subsection{$U(1)$-symmetric system}

Our motivations to study $U(1)$-symmetric systems are two-fold. The first obvious motivation is that we want to address the Hayden-Preskill recoverability and the AMPS problem in the presence of symmetries. The second, less obvious, motivation is to study the same set of questions for quantum systems which conserve total energy. 

Let us illustrate the second point. In the discussions from the previous section, we treated qubits on $B$ as coarse-grained degrees of freedom of a black hole with $S_{\text{BH}}=\log d_{\bar{B}}$. Instead, one might want to interpret $B$ as physical qubits on the boundary quantum system and the subspace $\bar{B}$ as a typical energy subspace of a black hole at finite temperature. Namely, if we consider the entangled AdS black hole, this amounts to assuming that $S_{E}=S_{\text{BH}} = \log d_{\bar{B}}$ while $\log d_{B}$ qubits are placed at UV. In this interpretation, our calculation would suggest that the recovery in the Hayden-Preskill thought experiment is not possible for a maximally entangled black hole at finite temperature. However, this conclusion is weird as a physical process akin to the Hayden-Preskill recovery has been recently found~\cite{Gao:2017aa}. It has been also argued that scrambling in a sense of decay of out-of-time order correlator is sufficient to perform recovery protocols even at finite temperature~\cite{Yoshida:2017aa}. Hence, something must be wrong in this interpretation. 

The error in the aforementioned argument can be traced back to the approximation of the black hole dynamics by Haar random unitary. Since Haar random unitary does not conserve energy, it brings quantum states on the input Hilbert space $AB$ to outside of the window of typical energy states, and hence recovering quantum states becomes harder. This observation motivates us to consider the AMPS problem and the Hayden-Preskill thought experiment by considering some version of Haar random unitary dynamics which captures physics of energy conserving systems.


One can mimic such a situation by considering $U(1)$-symmetric Haar random unitary where the total $U(1)$ charges are conserved. To be concrete, a $U(1)$-symmetric system can be modelled as a set of $n$ qubits where basis states can be expressed as an $n$-binary string and its total charge is defined as the number of $1$s:
\begin{align}
|i_{1},\ldots, i_{n}\rangle \qquad m = \sum_{k=1}^{n} i_{k} \qquad i_k =0,1.
\end{align}
We will consider the $U(1)$-symmetric Haar random unitary:
\begin{align}
U = \bigoplus_{m=0}^{n} U_{m} \qquad \mathcal{H} = \bigoplus_{m=0}^{n} \mathcal{H}_{m}
\end{align}
where $U_{m}$ is independently Haar random acting on each fixed-charge subspace $\mathcal{H}_{m}$. 

With energy conservation in mind, our goal was to consider a toy model of scrambling dynamics which mixes eigenstates with roughly equal energies. By viewing each fixed-charge subspace $\mathcal{H}_{m}$ as eigenstates with roughly equal energies, a $U(1)$-symmetric Haar random unitary can capture dynamics which throughly mixes quantum states from the small typical energy window. In a realistic quantum system, the Hilbert space structure does not decompose into a diagonal form in an exact manner. Here we hope to capture some salient feature of energy conserving systems in this simplified toy model.

\begin{figure}
\centering
\includegraphics[width=0.3\textwidth]{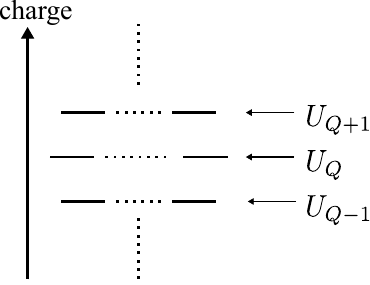} 
\caption{$U(1)$-symmetric Haar random unitary as a toy model of energy conserving dynamics.
\label{Penrose}
}
\end{figure}

\subsection{Haar integral}

Let us denote the local charge on $R$ by $m_{R}$ (\emph{i.e.} the total number of $1$'s in $R$). We will consider the cases where $A$ and $B$ have fixed charge $m_{A}$ and $m_{B}$ respectively. The quantum state of our interest is
\begin{align} 
|\Psi\rangle = \ \figbox{1.0}{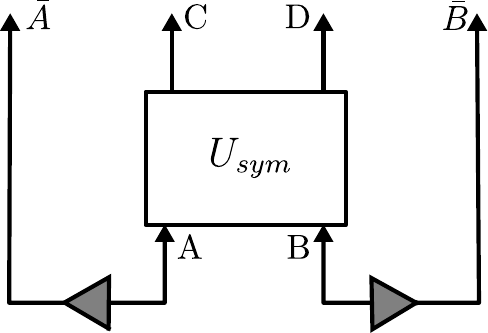}
\end{align}
where filled triangles represent normalized isometries onto fixed-charge subspaces. The input quantum states on $AB$ can be spanned in $\mathcal{H}_{in} = \mathcal{H}_{A}^{(m_{A})} \otimes \mathcal{H}_{B}^{(m_{B})}$ 
where $\mathcal{H}_{R}^{(m_{R})}$ represents a subspace of $\mathcal{H}_{R}$ with fixed charge $m_{R}$. Define $d_{\bar{A}}\equiv \dim \big(\mathcal{H}_{A}^{(m_{A})} \big)$ and $d_{\bar{B}}\equiv\dim \big(\mathcal{H}_{B}^{(m_{B})} \big)$. The input state in $\bar{A}A$ is given by
\begin{align}
\frac{1}{\sqrt{d_{\bar{A}}}}\sum_{k=1}^{d_{\bar{A}}} |k\rangle_{\bar{A}}|k_{\text{sym}}\rangle_{A}
\end{align}
where $|k_{\text{sym}}\rangle$ are states with fixed charges.  For instance, one may consider $|100\rangle, |010\rangle, |001\rangle$ for $n_{A}=3$ and $m_{A}=1$.

After a $U(1)$-symmetric unitary evolution, the total charge $m = m_{A} + m_{B}$ is conserved. The output Hilbert space is (assuming $n_{C}\geq m \geq n_{D}$)
\begin{align}
\mathcal{H}_{out} = \bigoplus_{Q=0}^{n_{D}} \mathcal{H}_{C}^{(m-Q)} \otimes \mathcal{H}_{D}^{(Q)}. \label{eq:Hilbert-space}
\end{align}
It is convenient to define
\begin{align}
d_{C}^{(m-Q)}=
\dim \big(\mathcal{H}_{C}^{(m-Q)} \big)  \qquad d_{D}^{(Q)} =\dim \big(\mathcal{H}_{D}^{(Q)} \big) \qquad  d = \dim\big( \mathcal{H}_{out} \big) = \sum_{Q=0}^{n_{D}} d_{C}^{(m-Q)}d_{D}^{(Q)}.
\end{align}
We have $d_{C}^{(m-Q)}={n_{C} \choose m-Q}$ and $d_{D}^{(Q)}={n_{D} \choose Q}$. 

Much of the analysis resembles the one in the previous section. The only complication is the treatment of delta functions when the Hilbert space does not have a direct product structure. The quantum state $|\Psi\rangle$ of the Hayden-Preskill thought experiment can be expressed as follows
\begin{align}
|\Psi\rangle = \frac{1}{\sqrt{d_{\bar{A}}d_{\bar{B}} }} U_{k\ell (s,t)}|k\rangle_{\bar{A}} |\ell\rangle_{\bar{B}} |(s,t)\rangle_{CD}
\end{align}
where $(s,t)$ indicates that summations over $(s,t)$ should be taken according to Eq.~\eqref{eq:Hilbert-space}. We find
\begin{align}
&\int dU \ \Tr \{ \rho_{C}^2 \} = \frac{1}{d_{\bar{A}}^2 d_{\bar{B}}^2} \int dU\ U_{k\ell (s,t)} U^{*}_{k \ell (s',t)} U_{k' \ell' (s',t')} U^{*}_{k' \ell' (s,t')} 
\end{align}
and a similar equation for $\Tr \{ \rho_{\bar{B}D}^2 \}$. 
In using the Haar formula, we need to apply delta functions to $(s,t)$. For instance, the first term of the Haar integral in $\Tr  \{ \rho_{C}^2 \}$ is 
\begin{align}
\frac{1}{d_{\bar{A}}^2 d_{\bar{B}}^2}\cdot (\text{number of $k, \ell, k',\ell'$})\cdot( \text{number of $(s,t)$ and $(s',t')$ with $s=s'$})\\
 = \frac{1}{d_{\bar{A}}^2 d_{\bar{B}}^2}\cdot d_{\bar{A}}^2 d_{\bar{B}}^2 \cdot \sum_{Q=0}^{n_{D}} d_{C}^{(m-Q)} \big( d_{D}^{(Q)} \big)^2 = \sum_{Q=0}^{n_{D}} d_{C}^{(m-Q)} \big( d_{D}^{(Q)} \big)^2.
\end{align}
It is convenient to define
\begin{align}
W_{C}\equiv \sum_{Q=0}^{n_{D}} \big( d_{C}^{(m-Q)} \big)^2 \cdot d_{D}^{(Q)}\qquad
W_{D}\equiv \sum_{Q=0}^{n_{D}} d_{C}^{(m-Q)} \cdot \big( d_{D}^{(Q)} \big)^2.
\end{align}
We find
\begin{align}
\int dU\ \Tr \{ \rho_{C}^2 \}  \approx  \frac{W_{D}}{d^2} + \frac{W_{C}}{d^2 d_{\bar{A}}d_{\bar{B}}} \qquad
\int dU\ \Tr \{ \rho_{\bar{B}D}^2 \} \approx \frac{W_{C}}{d^2 d_{\bar{B}}} + \frac{W_{D}}{d^2 d_{\bar{A}} }
\end{align}
after ignoring terms suppressed by $1/d$.

\subsection{Recoverability}

Let us find the criteria for recovery. If $n_{D}$ is large enough such that
\begin{align}
\frac{W_{D}}{d_{\bar{A}} } \gg \frac{W_{C}}{d_{\bar{B}}}, \label{eq:criteria}
\end{align}
we will have
\begin{align}
\Tr \{ \rho_{\bar{A}}^2 \} = \frac{1}{d_{\bar{A}}} \qquad \Tr \{ \rho_{C}^2 \}  \approx  \frac{W_{D}}{d^2} \qquad \Tr \{ \rho_{\bar{B}D}^2 \} \approx \frac{W_{D}}{d^2 d_{\bar{A}} }
\end{align}
implying $I^{(2)}(\bar{A},\bar{B}D)\approx 2\log_{2} d_{\bar{A}}$. Hence the recovery will be possible. As such, we need to find the condition on $n_{D}$ such that Eq.~\eqref{eq:criteria} holds. 

Let us write $W_{C}$ and $W_{D}$ as follows:
\begin{align}
W_{C} = \sum_{Q=0}^{n_{D}} W_{C}^{(Q)} \qquad W_{D} = \sum_{Q=0}^{n_{D}} W_{D}^{(Q)}
\end{align}
where
\begin{align}
W_{C}^{(Q)} \equiv {n_{C}\choose m-Q}^2   {n_{D}\choose Q} \qquad 
W_{D}^{(Q)} \equiv  {n_{C}\choose m-Q}  {n_{D}\choose Q}^2.
\end{align}
We look for the condition for the following inequality: 
\begin{align}
\frac{W_{D}^{(Q)}}{d_{\bar{A}}} \gg \frac{W_{C}^{(Q)}}{d_{\bar{B}}}.\label{eq:each}
\end{align}
By writing it down explicitly, we have 
\begin{align}
\frac{ {n_{D}\choose Q} }{d_{\bar{A}}} \gg \frac{{n_{C} \choose m - Q }}{  {n_{B} \choose m_{B} } }. \label{eq:Q_ineq}
\end{align}
When $Q \geq m_{A}$ and $n_{D} \geq n_{A}$, we have $m-Q\leq m_{B}$ and $n_{C}\leq n_{B}$. So, the RHS is smaller than unity. Hence it suffices to take 
\begin{align}
{n_{D}\choose Q} \gg d_{\bar{A}}
\end{align}
in order to satisfy Eq.~\eqref{eq:each}. 

While one cannot easily satisfy Eq.~\eqref{eq:each} by taking large $n_{D}$ for very small $Q$, contributions from such cases are negligibly small. In fact, for large $n_{C}$, we have 
\begin{align}
W_{C}^{(Q)} \simeq W_{C}^{(0)}{n_{D} \choose Q} \epsilon^{2Q} \qquad 
W_{D}^{(Q)} \simeq W_{D}^{(0)}{n_{D} \choose Q}^2 \epsilon^{Q}
\end{align}
with $\epsilon = \frac{p}{1-p}$. Since both $W_{C}^{(Q)}$ and $W_{D}^{(Q)}$ are (approximately) proportional to a binomial distribution and its square respectively,  contributions to $W_{C}$ and $W_{D}$ are dominated by $Q\sim O(n_{D})$. For such $Q$, it suffices to take $n_{D}\gg n_{A}$ in order for Eq.~\eqref{eq:each} to hold. Hence, we conclude that $n_{D}\gg n_{A}$ is sufficient for the recovery to be possible. We will present concrete recovery protocols in section~\ref{sec:recovery}. 

\subsection{Smoothness}

Next, let us compute the mutual information $I(C,D)$. For simplicity of discussion, we focus on the case where $d_{A}=1$, i.e. with no input state. In this case, $\rho_{CD}$ is a maximally mixed state with charge $m$:
\begin{align}
\rho_{CD} = \frac{1}{{ n \choose m}} \sum_{k=1}^{{ n \choose m}} |k_{\text{sym}}\rangle \langle k_{\text{sym}}|. 
\end{align}
This density matrix can be decomposed into a black diagonal form:
\begin{align}
\rho_{CD} = \sum_{Q=0}^{n_{D}} \text{Pr}(Q) {\rho_{CD}}^{(Q)} \qquad {\rho_{CD}}^{(Q)} = \sigma_{C}^{(m-Q)}  \otimes \sigma_{D}^{(Q)} 
\end{align}
where $\sigma_{C}^{(m-Q)}$ and $\sigma_{D}^{(Q)}$ represent maximally mixed states of charge $m-Q$ and $Q$ respectively. The probability weight is given by
\begin{align}
\text{Pr}(Q) = \frac{ {n_{D} \choose Q} { n_{C} \choose m-Q } }{{n \choose m}}. 
\end{align}
For large $n_{C}$, $\text{Pr}(Q)$ can be approximated by a binomial distribution with $p = m/n$. The mutual information is given by
\begin{align}
I(A,B) = - \sum_{Q=0}^{n_{D}} \text{Pr}(Q) \log \text{Pr}(Q) \simeq \frac{1}{2} \log n_{D}.
\end{align}
Since the variance of the binomial distribution is $\sim \sqrt{n_{D}}$, we can approximate it as a distribution over $\sqrt{n_{D}}$-level states which can be encoded in $\sim \frac{1}{2}\log n_{D}$ bits. The above argument suggests that these ``charge-bits'' are strongly correlated with those on $C$. However, as the block diagonal form suggests, the correlation between $C$ and $D$ is purely classical. 

\subsection{Physical interpretation}

Let us interpret the $U(1)$-symmetric Haar random unitary as an energy conserving dynamics. Then, the Hilbert space $B$ corresponds to physical qubits of a quantum system while $\bar{B}$ corresponds to the coarse-grained Hilbert space which is also the typical energy subspace at finite temperature. The black hole is maximally entangled with the early radiation in this interpretation; $S_{\text{BH}}=S_{E}=\log d_{\bar{B}}$. The Hawking radiation $D$ contains symmetric and non-symmetric modes which can be interpreted as soft and heavy modes respectively. 
When the input is symmetric with fixed charge, we found that the recovery is possible by collecting $O(1)$ qubits. 
However, as shown in appendix~\ref{appendix}, when the input is non-symmetric with variance in charge values, we found that the recovery requires $n_{D}$ to be extensive. With energy conserving systems in mind, this implies that input quantum states should be encoded in soft modes for recovery. We also see that the correlation between $C$ and $D$ is purely classical and results from charge conservation. With energy conserving systems in mind, it corresponds to correlations of heavy modes under energy conservation. Therefore, we interpret these heavy modes as Hawking quanta whereas soft modes are some entity responsible for the Hayden-Preskill recovery and scrambling dynamics. 

From the perspective of the AMPS problem, calculations in this section motivate us to consider two different kinds of operators on $D$. The off-diagonal operators are the ones which change the local charge in $D$. The partners of those off-diagonal operators can be identified in $C$ as operators which decrease and/or increase the total charge. This is due to classical (diagonal) correlations between $C$ and $D$. On the other hand, the diagonal operators are the ones which leave the total charge unchanged up to phases. Such operators can be explicitly written as follows:
\begin{align}
W = \sum_{n}e^{i \theta_{n}} |n\rangle\langle n|.
\end{align}
Unlike off-diagonal operators, the interior partners of diagonal operators $W$ cannot be found in $C$. Hence, the non-locality problem remains for partners of symmetric (soft) modes.


\section{Recovery via soft mode}\label{sec:recovery}

The goal of this section is to show that the Hayden-Preskill recovery can be performed via symmetric modes only. While we will study Haar random dynamics, we believe that similar conclusions hold for any ``scrambling'' systems in a sense of~\cite{Yoshida:2017aa} where scrambling is defined with respect to out-of-time order correlation functions.

We have discussed the recoverability in the Hayden-Preskill thought experiment without presenting explicit recovery protocols. The original work by Hayden and Preskill was essentially an existence proof of recovery protocols when the dynamics is given by Haar random unitary. Recently the author and Kitaev have constructed simple recovery protocols which work for any scrambling systems whose out-of-time order correlation functions decay~\cite{Yoshida:2017aa}. Similar recovery protocols can be applied to our $U(1)$-symmetric toy model. For simplicity of discussion, we will focus on a probabilistic recovery protocol. A deterministic protocol can be also constructed by following~\cite{Yoshida:2017aa}. Since the analysis in this section is a simple extension of the original work, we keep the presentation brief.

Consider the following quantum state:
\begin{align}
|\Phi\rangle \equiv \figbox{1.0}{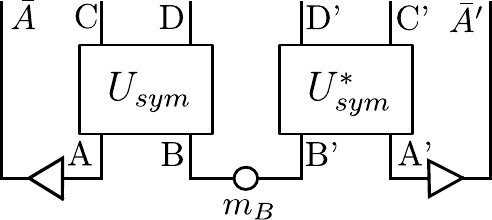}
\end{align}
where Bob prepared a particular quantum state on $A'\bar{A}'$ which is identical to the one on $A\bar{A}$, and applied the complex conjugate $U_{\text{sym}}^{*}$ on $B'$ and $A'$. The initial quantum state on $BB'$ is a maximally entangled symmetric state: $\frac{1}{\sqrt{d_{\bar{B}}}} \sum_{\ell=1}^{d_{\bar{B}}}  |\ell_{\text{sym}}\rangle_{B} \otimes |\ell_{\text{sym}}^{*}\rangle_{B'}$.
In the diagram, the unfilled dot with $m_{B}$ represents a normalized projection onto the subspace with total charge $m_{B}$. Bob has an access to $B'D$ (or $DD'C'\bar{A}'$), and his goal is to distill EPR pairs on $\bar{A}\bar{A}'$.

Bob's strategy is to perform a projection onto EPR pairs on $DD'$. Denoting the projector by $\Pi_{\text{EPR}}^{(DD')}$, the probability of measuring EPR pairs is 
\begin{align}
P_{\text{EPR}} = \langle \Phi | \Pi_{\text{EPR}}^{(DD')} | \Phi \rangle = \figbox{1.0}{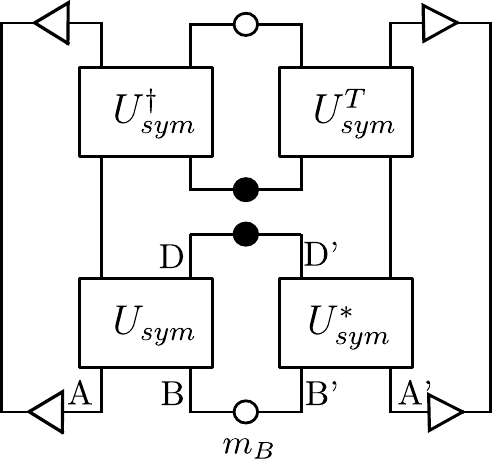}
\end{align}
where the filled dots in the middle represent projectors onto EPR pairs.
Recovery is successful if Bob can distill EPR pairs on $\bar{A}\bar{A}'$. Let us denote the fidelity of the distillation, conditioned on the measurement of EPR pairs on $DD'$, by $F_{\text{EPR}}$. The probability of measuring EPR pairs on both $DD'$ and $\bar{A}\bar{A}'$ is
\begin{align}
P_{\text{EPR}}F_{\text{EPR}} = \langle \Phi | \Pi_{\text{EPR}}^{(DD')} \Pi_{\text{EPR}}^{(\bar{A}\bar{A}')} | \Phi \rangle = \figbox{1.0}{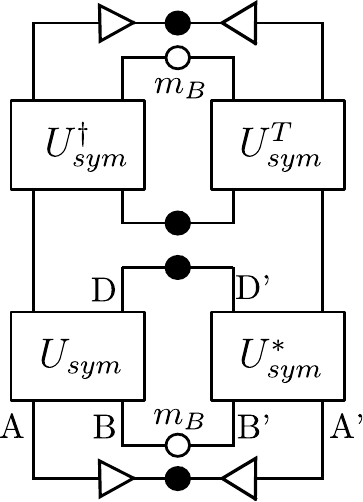} \ .
\end{align}

Both $P_{\text{EPR}}$ and $P_{\text{EPR}}F_{\text{EPR}}$ can be explicitly computed. From the above diagrams, we notice 
\begin{align}
P_{\text{EPR}} = \Tr (\rho_{\bar{B}D}^2) \frac{d_{\bar{B}}}{d_{D}} \qquad P_{\text{EPR}}F_{\text{EPR}} =\Tr (\rho_{C}^2) \frac{d_{\bar{B}}}{d_{D}d_{\bar{A}}}.
\end{align}
For $U(1)$-symmetric Haar random unitary, we obtain 
\begin{align}
P_{\text{EPR}} \simeq P_{\text{EPR}}F_{\text{EPR}} \simeq \frac{W_{D}d_{\bar{B}}}{d_{\bar{A}}d_{D}}
\end{align}
where subleading terms are suppressed for $n_{D}\gg n_{A}$. Hence, upon postselection, we have $F_{\text{EPR}}\simeq 1$ implying nearly perfect recovery.

The aforementioned recovery protocol can be modified to use only the symmetric mode. Namely, by applying a projection onto maximally entangled symmetric states on $DD'$, EPR pairs can be distilled on $\bar{A}\bar{A}'$. Let us denote the projector onto entangled states with charge $Q$ by $\Pi_{Q}^{(DD')}$. This projector is related to the EPR projector by
\begin{align}
\Pi_{\text{EPR}}^{(DD')} = \frac{1}{2^{n_{D}}} \sum_{Q=0}^{n_{D}} { n_{D}\choose Q } \Pi_{Q}^{(DD')}.
\end{align}
Let us denote the probability amplitude for measuring $\Pi_{Q}^{(DD')}$ by $P_{Q}$. For $U(1)$-symmetric Haar, we find 
\begin{align}
P_{Q} \simeq P_{Q}F_{\text{EPR}} \simeq { n_{C} \choose m-Q }{ n_{D} \choose Q}\frac{d_{\bar{B}}}{d^2d_{\bar{A}}}
\end{align}
for large $n_{D}$ which satisfies Eq.~\eqref{eq:Q_ineq}.
Hence, the Hayden-Preskill recovery can be carried out via symmetric modes.

\section{Construction of interior operator}\label{sec:mirror}

Finally, we discuss the construction of partner operators that would describe the interior mode in our toy model. The partner of non-symmetric (heavy) operators on $D$ can be easily constructed on the remaining black hole $C$ due to the classical correlations between $C$ and $D$. As such, we will focus on the partner of symmetric (soft) operators. 

The main result of this section is the observation that construction of interior partner operators can be interpreted as the Hayden-Preskill recovery problem in disguise. This observation enables us to show that the black hole interior modes are robust against perturbations on the early radiation due to scrambling dynamics. Namely, even if almost all the qubits, except a few, in the early radiation are damaged, interior partner operators can be still constructed. 

We assume $d_{A}=1$ although our construction works well for cases with $d_{A}>1$ too. 

\subsection{Interior from Hayden-Preskill}

Interestingly, reconstruction of the interior operators can be performed by using a procedure similar to the Hayden-Preskill recovery. Let us begin by defining what we mean by interior partner operators. The quantum state of our interest is as follows:
\begin{align}
|\Psi\rangle \ = \ \figbox{1.0}{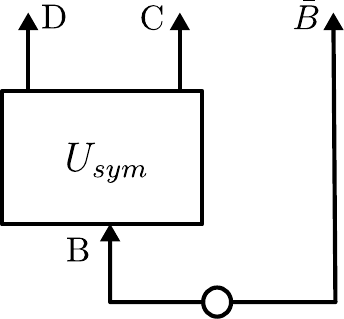}\ 
\end{align}
where $D$ is the outgoing mode and $C$ is the remaining black hole.
Our task is the following; given a symmetric (diagonal) operator $O_{D}$, find the partner operator $V_{C\bar{B}}$ such that 
\begin{align}
(O_{D}\otimes I_{C\bar{B}})|\Psi\rangle \simeq (I_{D}\otimes V_{C\bar{B}})|\Psi\rangle.
\end{align}
Graphically the above equation reads
\begin{align}
\figbox{1.0}{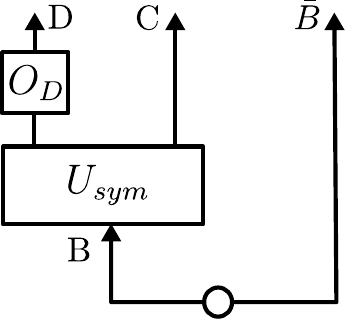}\ \simeq \ \figbox{1.0}{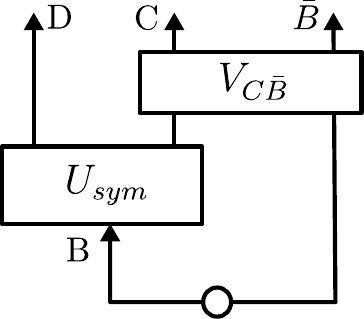} \ . \label{eq:equation-diagram}
\end{align}
Note that the existence of a partner operator is guaranteed. The question concerns how to write it down. 

To address this problem, it is convenient to interpret the above quantum state $|\Psi\rangle$ as a map from $D$ to $C\overline{B}$ where (symmetric) quantum states on $D$ are encoded into $C\bar{B}$. To make this interpretation more concrete, let us deform the diagrams in Eq.~\eqref{eq:equation-diagram} in the following manner:
\begin{align}
\figbox{1.0}{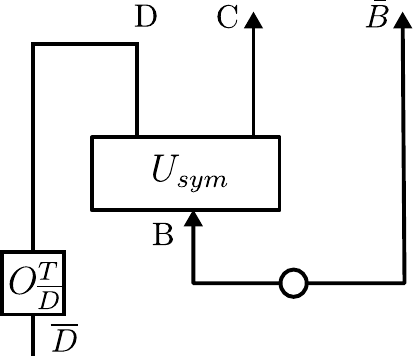} \ \simeq \ \figbox{1.0}{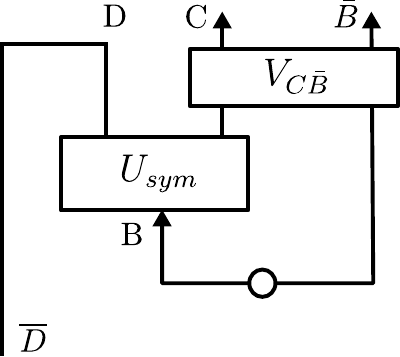}
\end{align}
Here the original diagrams had three upward ``output'' legs whereas the deformed graphs use the $D$-index as an ``input'' by bending it downward. The new diagrammatic equation represents a map from $O_{\overline{D}}^T$ to a partner operator $V_{C\overline{B}}$. The map itself can be explicitly expressed as follows:
\begin{align}
\overline{D} \rightarrow C\overline{B} \ : \ \figbox{1.0}{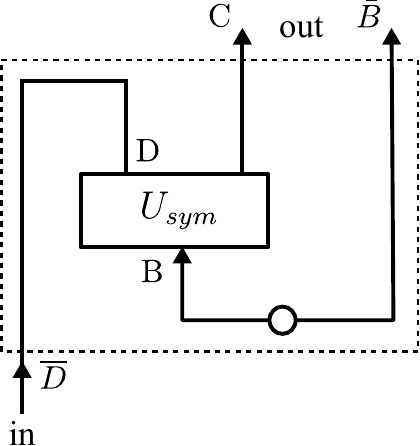}.
\end{align}
This is an isometric embedding (preserving inner products) from a smaller Hilbert space $\overline{D}$ onto a larger one $C \overline{B}$. It is worth recalling that $\overline{B}$ and $C$ correspond to the early radiation and the remaining black hole respectively. We can associate a physical process to the deformed diagram by reading it from the bottom to the top. Namely, it begins with a system consisting of an arbitrary initial state on $\overline{D}$ and a maximally entangled symmetric state on $B\overline{B}$. Then the system evolves by $U_{\text{sym}}$, and is projected onto the EPR pair on $\overline{D}D$ to obtain an output wavefunction on $C\overline{B}$.

One can further simplify the above map (and make the relation to the Hayden-Preskill problem more explicit) by rotating the box of the unitary operator $U_{\text{sym}}$ by 180 degrees. Carefully redrawing the diagram, we obtain the following(s):
\begin{align}
\figbox{1.0}{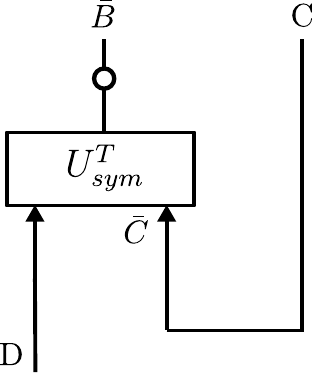} \ = \ \figbox{1.0}{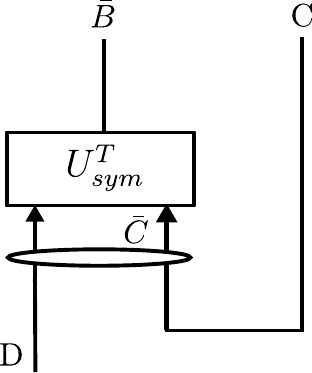}
\end{align}
Here $U_{\text{sym}}^T$ represents the transpose of $U_{\text{sym}}$, which results from flipping the diagram upside down. The circle on $\overline{B}$ represent a projection onto a symmetric subspace with fixed charge. An elongated circle also represents a projection onto a symmetric subspace with fixed charge. Note that the projector does not factor on $D \otimes \bar{C}$. Also note that the projector commutes with symmetric operator $O_{D}$ and $U_{\text{sym}}^T$.  


Our task is to reconstruct a partner of $O_D^T$ on $\bar{B}C$. 
Of course, it is possible to find a partner operator on the early radiation $\bar{B}$ by simply time-evolving $O_{D}^T$ by $U_{\text{sym}}^T$. However, the construction of a partner operator $V_{C\bar{B}}$ is not unique since since $\bar{B}C$ is larger than $D$. The non-uniqueness of the interior partner operator is closely related to the fact that the above quantum state $|\Psi\rangle$ can be interpreted as a quantum error-correcting code where (symmetric) quantum states on $D$ are encoded into $C\bar{B}$. 

Here we want to find an alternative representation which involves as few qubits on $\bar{B}$ as possible. A key observation is that, in the above diagram, the outgoing mode $D$ can be interpreted as an input Hilbert space for the Hayden-Preskill thought experiment where the unitary evolution is given by $U_{\text{sym}}^T$. To be explicit, let $\bar{B}_{0}$ be some small subsystem of the early radiation $\bar{B}$ which contains $n_{\bar{B}_{0}}\sim O(1)$  qubits such that $n_{\bar{B}_{0}} \gg n_{D}$. Let $\bar{B}_{1}$ be the complement of $\bar{B}_{0}$ in $\bar{B}$. The recoverability in the Hayden-Preskill thought experiment for symmetric modes implies that there exists a partner operator $V_{\bar{B}_{0}C}$ supported on $\bar{B}_{0}C$: 
\begin{align}
\figbox{1.0}{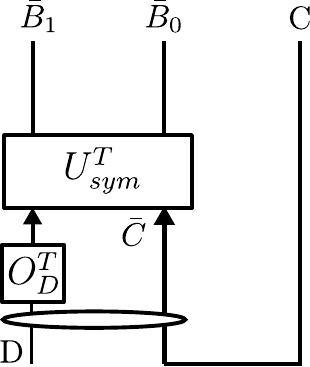} \  = \ \figbox{1.0}{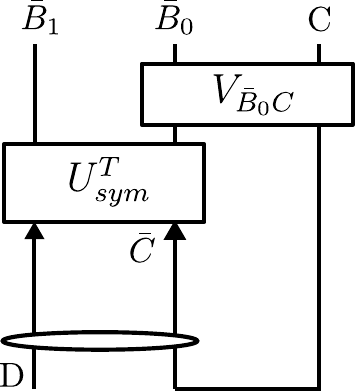}\ .
\end{align}
While we assumed $d_{A}=1$, the above procedure works for $d_{A}>1$ cases since $\bar{B}_{0}$ can be chosen arbitrarily. 

We have observed that the reconstruction of the interior operators is essentially the Hayden-Preskill thought experiment. This suggests the following result:
\begin{itemize}
\item While the partner operator of the outgoing soft mode $D$ cannot be found in the remaining black hole $C$, adding a few extra qubits $\bar{B}_{0}$ from the early radiation $\bar{B}$ to $C$ is enough to construct the interior operator. Due to the scrambling nature of $U_{\text{sym}}^T$, one may choose any set of a few qubits $\bar{B}_{0}$ in order to construct the interior operator as long as out-of-time order correlation functions between $D$ and $\bar{B}_{0}$ decay. 
\end{itemize}

\subsection{Fault-tolerance}

From the outside quantum mechanical viewpoint, the AMPS problem (or the non-locality problem) originates from the fact that $I(C,D)$ is small and the partner operator of $D$ cannot be constructed in the remaining black hole $C$. The construction of interior operators via the Hayden-Preskill recovery protocols enables us to construct a partner operator on $C\bar{B}_{0}$ where $\bar{B}_{0}$ is a small subsystem with $|\bar{B}_{0}|\gtrapprox |D|$. While we still need to include a few extra qubits from the early radiation $\bar{B}$, the construction is ``almost'' inside the remaining black hole $C$ !

In fact, this observation sheds a new light on questions concerning the robustness of the black hole interior against perturbations on the early radiation.
Some previous works attempted to resolve the AMPS puzzle by using the concept of quantum circuit complexity by drawing distinctions between simple and complex quantum operations~\cite{Harlow:2013aa, Maldacena13}.
Namely, it has been argued that action of simple quantum operations on the early radiation should not disturb the black hole interior operators~\cite{Maldacena13}.
Heuristic explanations for such fault-tolerant ``encoding'' of interior operators have been presented by using an analogy with quantum error-correcting code~\cite{Almheiri:2015ac, Verlinde:2013aa, Almheiri2018, Verlinde13a, Verlinde13b, Goel18}. However concrete physical origin of robustness of the black hole interior mode was not discussed from the outside quantum mechanical perspective.

Our construction of interior operators suggests that robustness of the black hole interior arises from scrambling dynamics.  
The essential point is that the small subsystem $\bar{B}_{0}$ in $\bar{B}$ can be chosen in an arbitrary manner as long as the out-of-time order correlation functions between $\bar{B}_{0}$ and $D$ decay~\cite{Hosur:2015ylk, Yoshida:2017aa}.
In particular, let us consider a scenario where some large perturbations are added on the early radiation $\bar{B}$ which would damage all the qubits on $\bar{B}_{1}$ ($n - \log |\bar{B}_0|$ qubits).
The construction of the interior operator is immune to such a drastic error since it does not involve any qubits from $\bar{B}_{1}$.
In this sense, our construction is naturally fault-tolerant against perturbations on the early radiation $\bar{B}$. 
This unusual robustness of the encoding of interior operators results from the very fact that they can be constructed via the Hayden-Preskill recovery protocols.

The remaining question concerns how to write down the interior operators explicitly. On a formal level, if the time evolution $U_{\text{sym}}^T$ is a scrambling unitary with the decay of out-of-time order correlation functions, the interior operators can be explicitly constructed by running the recovery protocol proposed in~\cite{Yoshida:2017aa}. We are currently working on specific quantum systems and observe that the concept of operator growth perspective plays important roles in the construction. This will be presented elsewhere. 

%
%

\section{Discussions}\label{sec:discussion}

In this paper, we addressed the tension between the smoothness of the horizon and the recoverability in the Hayden-Preskill thought experiment by using a toy model with energy conservation. Within the validity of the toy model, our calculation suggests that the Hawking radiation corresponds to heavy modes whereas the Hayden-Preskill thought experiment must concern soft modes only. The correlation between the remaining black hole and the outgoing radiation is found to be classical. The classical correlation remains due to the energy conservation and due to the fact that the black hole in our toy model is entangled with the early radiation only through soft modes. Our toy model suggests that the off-diagonal correlation decoheres since the phases of heavy modes are scrambled by chaotic dynamics in soft modes. Finally, we observed that the procedure of reconstructing the soft part of the infalling mode can be interpreted as the Hayden-Preskill recovery protocol. As such, while the description of the infalling mode may require the early radiation, only a few extra qubits will be sufficient. In the reminder of the paper, we present discussions on relevant topics. 

\subsection{AMPS puzzle and scrambling}

Our construction of interior operators via the Hayden-Preskill recovery phenomenon sheds a new light on the AMPS puzzle through the lens of quantum information scrambling. For simplicity of discussion, let us consider a maximally entangled black hole at infinite temperature which is represented by $n$ copies of EPR pairs. To recap briefly, the AMPS puzzle concerns an apparent tension between descriptions by the infalling observer and the outside observer. From the perspective of the outside observer, the outgoing mode $D$ is entangled with some degrees of freedom in the early radiation $\bar{B}$. From the perspective of the infalling observer, the same outgoing mode $D$ must be entangled with some interior mode, leading to violation of the monogamy of quantum entanglement.

\begin{figure}
\centering
\includegraphics[width=0.27\textwidth]{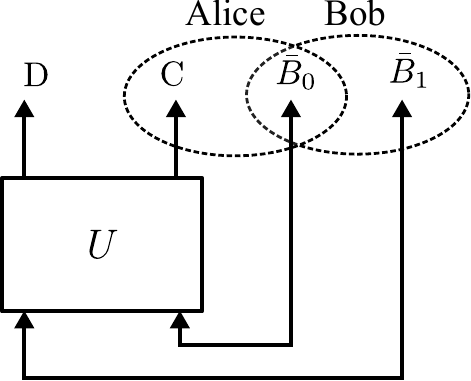} 
\caption{The AMPS puzzle and scrambling. Alice, an infalling observer, can distill an EPR pair by accessing $\bar{B}_{0}C$ while Bob, an outside observer, can distill an EPR pair by accessing $\bar{B}=\bar{B}_{0}\bar{B}_{1}$. Note that $\bar{B}_{0}$ can be any subsystem of the early radiation $\bar{B}$ as long as $|\bar{B}_{0}|\gtrapprox |D|$ due to scrambling property of $U$.
}
\label{fig-fight}
\end{figure}

The intriguing lesson from our construction of interior operators is that whoever possesses a tiny portion $\bar{B}_{0}$ of the early radiation $\bar{B}$ will be able to distill a qubit which is entangled with the outgoing mode $D$. Namely, if $\bar{B}_{0}$ is included to $C$ as degrees of freedom which the infalling observer can touch, she can distill an EPR pair between $D$ and $\bar{B}_{0}C$. On the other hand, if $\bar{B}_{0}$ is left untouched by the infalling observer, the outgoing mode $D$ is entangled with $\bar{B}=\bar{B}_{0}\bar{B}_{1}$ and the outside observer can distill an EPR pair between $D$ and $\bar{B}$. See Fig.~\ref{fig-fight} for schematic illustration. In fact, these statements can be made quantitative. Decay of out-of-time order correlation functions implies that $I^{(2)}(D,\bar{B}_{0}C)$ is nearly maximal~\cite{Hosur:2015ylk}. Since $I^{(2)}(D,\bar{B}_{0}C)+I^{(2)}(D,\bar{B}_{1})=2 S_{D}^{(2)}$, this suggests that $I^{(2)}(D,\bar{B}_{1})$ is close to zero. Namely, the infalling observer may reconstruct a partner operator on $\bar{B}_{0}C$ while the outside observer cannot reconstruct it on $\bar{B}_{1}$. 

It is worth emphasizing that $\bar{B}_{0}$ can be any subsystem of $\bar{B}$ as long as $|\bar{B}_{0}|\gtrapprox |D|$. Hence, from the perspective of the infalling observer, it is rather easy to disentangle the outgoing mode $D$ from the early radiation $\bar{B}$ as she needs to touch only a few qubits in $\bar{B}$. In fact, it turns out that quite generic perturbations to the black hole by the infalling observer will disentangle the outgoing mode $D$ from the early radiation $\bar{B}$ \emph{without} ever accessing the early radiation $\bar{B}$. In an accompanying paper, based on this disentangling phenomena, we will propose a possible resolution of the AMPS problem in a manner which is free from the non-locality problem and the state-dependence problem~\cite{Yoshida:2019}.

\subsection{Soft modes $\simeq$ codewords}

Our toy model crucially relies on the assumption that there is a clear separation between heavy and soft modes. The key insight behind this simplification is that a few thermodynamic quantities, such as energy and charge, determine the underlying classical geometry. This is essentially the statement of the no hair theorem of classical black holes. However, there are many black hole micro-states which are consistent with the given classical geometry. The soft mode, discussed in the toy model, aims to capture all of these extremely low energy degrees of freedom. While it is unclear to us to what extent this toy model can capture the actual physics, it is concrete and simple enough to make theoretically verifiable predictions. 

One interesting point is that a simple toy model with $U(1)$ symmetry can be interpreted as a energy conserving system and naturally gives rise to heavy and soft modes. In a more generic setting, we may imagine an approximate decomposition of the full Hilbert space into a block diagonal form:
\begin{align}
\mathcal{H} \approx \bigoplus_{E,Q} \mathcal{H}_{E,Q}
\end{align}
where $E$ represents the energy and $Q$ represents the charge, angular momentum and other relevant macroscopic quantities. Each subspace $\mathcal{H}_{E,Q}$ defines a Hilbert space for the classical geometry determined by a set of $E,Q$. Heavy operators correspond to those which moves between different subspaces whereas soft modes correspond to degrees of freedom inside $\mathcal{H}_{E,Q}$. In realistic situations, such a decomposition into the block diagonal form will be an approximate one. Also there are ambiguities on which degrees of freedom should be treated as soft modes. For instance, depending on the problems of interest and energy/time scales as well as dimensionality, matter on the bulk may be considered as either soft or hard mode. Our toy model aims to capture the idealistic limit where the decomposition becomes exact with sharp distinction between heavy and soft modes. 

It is worth recalling that separation of soft and heavy modes plays important roles in a number of problems in quantum gravity. To add a more speculative comment, the quantum error-correcting property in the AdS/CFT correspondence is a manifestation of such separation of energy scales~\cite{Almheiri:2015aa, Pastawski15b}. In this interpretation, the geometry $(E,Q)$ determines the codeword subspace $\mathcal{H}_{E,Q}$ while the low energy modes correspond to different codeword states in a quantum error-correcting code determined by $(E,Q)$. The ``errors'' in this quantum error-correcting code $\mathcal{H}_{E,Q}$ are heavy operators which moves the system to the outside of the codeword subspace $\mathcal{H}_{E,Q}$. In this sense, our toy model is an attempt to apply the idea of quantum error-correction to dynamical problems in quantum gravity. Hence, we believe that our approach of using the $U(1)$-symmetric toy model, despite being very simple, is applicable to a wide variety of interesting questions in quantum gravity. 

At this moment, however, it is unclear to us how the black hole evaporates and eventually gets entangled only through soft modes. On one hand, if we assume that the underlying geometry changes adiabatically during evaporation, then it is reasonable to assume that the fluctuation of energy, or more generically heavy modes, is greatly suppressed. On the other hand, as is clear from the calculation of $I(C,D)$ in our toy model, the process of emitting the Hawking radiation does introduce fluctuations whose energy scale is much larger than soft modes by definition. Hence, in order for our toy model to be applicable, there needs to be some physical mechanism to suppress the energy variance in a dynamical manner. 

\subsection{Factorization of Hilbert space}

Throughout the paper we used a toy model that represents a black hole as a system of qubits. This is a drastic simplification building on two non-trivial assumptions. First, it is assumed that the black hole Hilbert space is discrete. Second, it is assumed that the Hilbert space of the black hole is factorizable. The first assumption is relatively well justified as it stems from the finiteness of the black hole entropy. On the other hand, the second assumption is incorrect in a strict sense. 

Whether the Hayden-Preskill recovery (as well as the AMPS argument) is applicable to systems with non-factorizable Hilbert spaces is an interesting problem. We envision that this question can be ultimately answered by using operator algebraic approaches, defining Hilbert spaces from operators instead of starting from a given Hilbert space. For such an extension, we would need to define entanglement and recoverability in operator algebraic languages. While providing a full-fledged answer to this question is clearly beyond the scope of this paper, our analysis may be viewed as a first step toward this question. In this paper, we discussed the Hayden-Preskill recovery problem in the presence of symmetries. A symmetric subspace, even if it is embedded on a factorizable Hilbert space, is known to be non-factorizable. We studied conditions under which the recovery is possible and saw that distinction between symmetric and non-symmetric operators is crucial. This observation hints that the criteria on recoverability may be stated purely in terms of operators without direct use of entanglement and the structure of a given Hilbert space.

\section*{Acknowledgment}

I am grateful to Nick Hunter-Jones for collaboration during the initial phase of the project. 
I would like to thank Ahmed Almheiri, Nick Hunter-Jones, Naritaka Oshita, John Preskill, Brian Swingle and Herman Verlinde for useful discussions on related topics. Research at Perimeter Institute is supported by the Government of Canada through Industry Canada and by the Province of Ontario through the Ministry of Research and Innovation.

\appendix

\section{U(1)-Symmetric Hayden-Preskill; non-symmetric inputs}\label{appendix}

Here we consider the Hayden-Preskill thought experiment with $U(1)$-symmetric time-evolution when an input quantum state is non-symmetric. To be specific, we will consider the input states which are superpositions of the following states with different charges:
\begin{align}
|\tilde{j}\rangle = |1\cdots 1 0 \cdots \rangle \qquad j = 0,\cdots, Q
\end{align}
where the first $j$ entries are $1$'s. So we have $n_{A}=Q$ and $n_{B}=n-Q$. The quantum state of our interest is 
\begin{align}
|\Psi\rangle = \ \figbox{1.0}{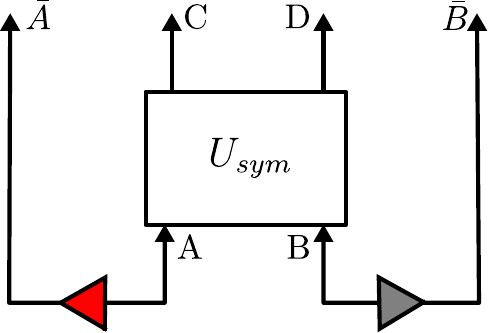}
\end{align}
where the red triangles on $A\bar{A}$ corresponds to $\frac{1}{\sqrt{Q+1}} \sum_{j=0}^{Q} |j\rangle_{\bar{A}} \otimes |\tilde{j}\rangle_{A}$. Let us denote the Hilbert space with total charge $m_{B}+a$ by $\mathcal{H}_{a}$ and the Haar random unitary acting on it by $U_{a}$. By directly computing $\Tr\{ \rho_{C}^2 \}$ and $\Tr\{ \rho_{\bar{B}D}^2 \}$, one can show that the reconstruction does not work. Namely we have $I^{(2)}(A,\bar{B}D), I^{(2)}(A,C) \approx \log (Q+1)$ when $D\sim O(1)$. This suggests that diagonal information about the total charge cannot be reconstructed from either $C$ or $\bar{B}D$.


\providecommand{\href}[2]{#2}\begingroup\raggedright\endgroup


\begin{thebibliography}{10}

\bibitem{Hawking75}
S.~Hawking, ``Particle creation by black holes,'' {\em Commun. Math. Phys.}
  {\bfseries 43} (1975) 199--220.

\bibitem{Page93}
D.~N. Page, ``Average entropy of a subsystem,'' {\em Phys. Rev. Lett.}
  {\bfseries 71} (1993) 1291--1294.

\bibitem{Page:1993aa}
D.~N. Page, ``Information in black hole radiation,''
  \href{http://dx.doi.org/10.1103/PhysRevLett.71.3743}{{\em Phys. Rev. Lett.}
  {\bfseries 71} (1993) 3743--3746}.

\bibitem{Almheiri13}
A.~Almheiri, D.~Marolf, J.~Polchinski, and J.~Sully, ``Black holes:
  complementarity or firewalls?,''
  \href{http://dx.doi.org/10.1007/JHEP02(2013)062}{{\em JHEP} {\bfseries 02}
  (2013) 062}.

\bibitem{Hayden07}
P.~Hayden and J.~Preskill, ``Black holes as mirrors: quantum information in
  random subsystems,'' {\em JHEP} {\bfseries 09} (2007) 120.

\bibitem{Braunstein:2013aa}
S.~L. Braunstein, S.~Pirandola, and K.~{\.Z}yczkowski, ``Better late than
  never: Information retrieval from black holes,''
  \href{http://dx.doi.org/10.1103/PhysRevLett.110.101301}{{\em Phys. Rev. Lett}
  {\bfseries 110} (2013) 101301}.

\bibitem{Arkani-Hamed:2007aa}
N.~Arkani-Hamed, Lubo{\v s}Motl, A.~Nicolis, and C.~Vafa, ``The string
  landscape, black holes and gravity as the weakest force,'' {\em JHEP}
  {\bfseries 06} (2007) 060.

\bibitem{Khemani:2018aa}
V.~Khemani, A.~Vishwanath, and D.~A. Huse, ``Operator spreading and the
  emergence of dissipative hydrodynamics under unitary evolution with
  conservation laws,'' \href{http://dx.doi.org/10.1103/PhysRevX.8.031057}{{\em
  Phys. Rev. X} {\bfseries 8} (2018) 031057}.

\bibitem{Rakovszky:2018aa}
T.~Rakovszky, F.~Pollmann, and C.~W. von Keyserlingk, ``Diffusive hydrodynamics
  of out-of-time-ordered correlators with charge conservation,''
  \href{http://dx.doi.org/10.1103/PhysRevX.8.031058}{{\em Phys. Rev. X}
  {\bfseries 8} (2018) 031058}.

\bibitem{Hooft:1987aa}
G.~'t~Hooft, ``Strings from gravity,'' {\em Physica Scripta} {\bfseries 15}
  (1987) 143.

\bibitem{Strominger:2014aa}
A.~Strominger, ``On bms invariance of gravitational scattering,''
  \href{http://dx.doi.org/10.1007/JHEP07(2014)152}{{\em JHEP} {\bfseries 7}
  (2014) 152}.

\bibitem{Yoshida:2017aa}
B.~Yoshida and A.~Kitaev, ``Efficient decoding for the hayden-preskill
  protocol,'' \href{http://arxiv.org/abs/arXiv:1710.03363}{{\ttfamily
  arXiv:1710.03363}}.

\bibitem{Almheiri13b}
A.~Almheiri, D.~Marolf, J.~Polchinski, D.~Stanford, and J.~Sully, ``An apologia
  for firewalls,'' \href{http://dx.doi.org/10.1007/JHEP09(2013)018}{{\em JHEP}
  {\bfseries 09} (2013) 018}.

\bibitem{Papadodimas:2013aa}
K.~Papadodimas and S.~Raju, ``An infalling observer in ads/cft,''
  \href{http://dx.doi.org/10.1007/JHEP10(2013)212}{{\em JHEP} {\bfseries 10}
  (2013) 212}.

\bibitem{Maldacena13}
J.~Maldacena and L.~Susskind, ``Cool horizons for entangled black holes,''
  \href{http://dx.doi.org/10.1002/prop.201300020}{{\em Fortsch. Phys.}
  {\bfseries 61} (2013) 781--811}.

\bibitem{Susskind13}
L.~Susskind, ``New concepts for old black holes.'' Arxiv:1311.3335.

\bibitem{Bousso:2013ab}
R.~Bousso, ``Complementarity is not enough,''
  \href{http://dx.doi.org/10.1103/PhysRevD.87.124023}{{\em Phys. Rev. D}
  {\bfseries 87} (2013) 124023--}.

\bibitem{Marolf:2013aa}
D.~Marolf and J.~Polchinski, ``Gauge-gravity duality and the black hole
  interior,'' \href{http://dx.doi.org/10.1103/PhysRevLett.111.171301}{{\em
  Phys. Rev. Lett.} {\bfseries 111} (2013) 171301--}.

\bibitem{Bousso:2013aa}
R.~Bousso, ``Firewalls from double purity,''
  \href{http://dx.doi.org/10.1103/PhysRevD.88.084035}{{\em Phys. Rev. D}
  {\bfseries 88} (2013) 084035--}.

\bibitem{Giddings:2013aa}
S.~B. Giddings, ``Nonviolent nonlocality,''
  \href{http://dx.doi.org/10.1103/PhysRevD.88.064023}{{\em Phys. Rev. D}
  {\bfseries 88} (2013) 064023--}.

\bibitem{Harlow:2016ab}
D.~Harlow, ``Jerusalem lectures on black holes and quantum information,''
  \href{http://dx.doi.org/10.1103/RevModPhys.88.015002}{{\em Rev. Mod. Phys.}
  {\bfseries 88} (2016) 015002--}.

\bibitem{Bousso:2014aa}
R.~Bousso, ``Violations of the equivalence principle by a nonlocally
  reconstructed vacuum at the black hole horizon,''
  \href{http://dx.doi.org/10.1103/PhysRevLett.112.041102}{{\em Phys. Rev.
  Lett.} {\bfseries 112} (2014) 041102--}.

\bibitem{Marolf:2016aa}
D.~Marolf and J.~Polchinski, ``Violations of the born rule in cool
  state-dependent horizons,''
  \href{http://dx.doi.org/10.1007/JHEP01(2016)008}{{\em JHEP} {\bfseries 1}
  (2016) 8}.

\bibitem{Harlow:2014aa}
D.~Harlow, ``Aspects of the papadodimas-raju proposal for the black hole
  interior,'' \href{http://dx.doi.org/10.1007/JHEP11(2014)055}{{\em JHEP}
  {\bfseries 11} (2014) 55}.

\bibitem{Papadodimas:2014aa}
K.~Papadodimas and S.~Raju, ``Black hole interior in the holographic
  correspondence and the information paradox,''
  \href{http://dx.doi.org/10.1103/PhysRevLett.112.051301}{{\em Phys. Rev.
  Lett.} {\bfseries 112} (2014) 051301--}.

\bibitem{Papadodimas:2014ab}
K.~Papadodimas and S.~Raju, ``State-dependent bulk-boundary maps and black hole
  complementarity,'' \href{http://dx.doi.org/10.1103/PhysRevD.89.086010}{{\em
  Phys. Rev. D} {\bfseries 89} (2014) 086010--}.

\bibitem{Raju:2017aa}
S.~Raju, ``Smooth causal patches for ads black holes,''
  \href{http://dx.doi.org/10.1103/PhysRevD.95.126002}{{\em Phys. Rev. D}
  {\bfseries 95} (2017) 126002--}.

\bibitem{Jafferis17}
D.~L. Jafferis, ``Bulk reconstruction and the hartle-hawking wavefunction.''
  Arxiv:1703.01519.

\bibitem{Yoshida:2019}
B.~Yoshida, ``Firewalls vs. scrambling,''
  \href{http://arxiv.org/abs/arXiv:1902.09763}{{\ttfamily arXiv:1902.09763}}.

\bibitem{Hosur:2015ylk}
P.~Hosur, X.-L. Qi, D.~A. Roberts, and B.~Yoshida, ``{Chaos in quantum
  channels},'' \href{http://dx.doi.org/10.1007/JHEP02(2016)004}{{\em JHEP}
  {\bfseries 02} (2016) 004}.

\bibitem{Nomura18}
Y.~Nomura, ``Reanalyzing an evaporating black hole,''
  \href{http://arxiv.org/abs/arXiv:1810.09453}{{\ttfamily arXiv:1810.09453}}.

\bibitem{Verlinde:2013aa}
E.~Verlinde and H.~Verlinde, ``Black hole entanglement and quantum error
  correction,'' \href{http://dx.doi.org/10.1007/JHEP10(2013)107}{{\em JHEP}
  {\bfseries 10} (2013) 107}.

\bibitem{Yoshida:2019aa}
B.~Yoshida and N.~Y. Yao, ``Disentangling scrambling and decoherence via
  quantum teleportation,''
  \href{http://dx.doi.org/10.1103/PhysRevX.9.011006}{{\em Phys. Rev. X}
  {\bfseries 9} (2019) 011006--}.

\bibitem{Gao:2017aa}
P.~Gao, D.~L. Jafferis, and A.~C. Wall, ``Traversable wormholes via a double
  trace deformation,'' \href{http://dx.doi.org/10.1007/JHEP12(2017)151}{{\em
  JHEP} {\bfseries 12} (2017) 151}.

\bibitem{Harlow:2013aa}
D.~Harlow and P.~Hayden, ``Quantum computation vs. firewalls,''
  \href{http://dx.doi.org/10.1007/JHEP06(2013)085}{{\em JHEP} {\bfseries 6}
  (2013) 85}.

\bibitem{Almheiri:2015ac}
A.~Almheiri, X.~Dong, and D.~Harlow, ``Bulk locality and quantum error
  correction in ads/cft,''
  \href{http://dx.doi.org/10.1007/JHEP04(2015)163}{{\em JHEP} {\bfseries 4}
  (2015) 163}.

\bibitem{Almheiri2018}
A.~Almheiri, ``Holographic quantum error correction and the projected black
  hole interior,'' \href{http://arxiv.org/abs/arXiv:1810.02055}{{\ttfamily
  arXiv:1810.02055}}.

\bibitem{Verlinde13a}
E.~Verlinde and H.~Verlinde, ``Behind the horizon in ads/cft.''
  Arxiv:1311.1137.

\bibitem{Verlinde13b}
E.~Verlinde and H.~Verlinde, ``Passing through the firewall.'' Arxiv:1306.0515.

\bibitem{Goel18}
A.~Goel, H.~T. Lam, G.~J. Turiaci, and H.~Verlinde, ``Expanding the black hole
  interior: Partially entangled thermal states in syk.'' Arxiv:1807.03916.

\bibitem{Almheiri:2015aa}
A.~Almheiri, X.~Dong, and D.~Harlow, ``Bulk locality and quantum error
  correction in ads/cft,''
  \href{http://dx.doi.org/10.1007/JHEP04(2015)163}{{\em JHEP} {\bfseries 04}
  (2015) 163}.

\bibitem{Pastawski15b}
F.~Pastawski, B.~Yoshida, D.~Harlow, and J.~Preskill, ``Holographic quantum
  error-correcting codes: toy models for the bulk/boundary correspondence,''
  \href{http://dx.doi.org/10.1007/JHEP06(2015)149}{{\em JHEP} {\bfseries 06}
  (2015) 149}.

\end{thebibliography}
%
\end{document}